\documentclass[letter]{aa} % for the letters 
\usepackage{graphicx}
\usepackage{txfonts}
\begin{document}

   \title{Frequency-dependent tidal dissipation in a viscoelastic Saturnian core and expansion of Mimas' semi-major axis}
 \titlerunning{Saturnian dissipation by frequency dependent core}
  % \subtitle{I. Overviewing the $\kappa$-mechanism}

   \author{D. Shoji
         % \inst{1}
          \and
          H. Hussmann%\inst{1}%\fnmsep\thanks{Just to show the usage
       %  of the elements in the author field}
          }

   \institute{German Aerospace Center(DLR), Institute of Planetary Research, Rutherfordstr. 2 12489 Berlin, Germany\\
              \email{[Daigo.Shoji; Hauke.Hussmann]@dlr.de}
         %\and
          %   University of Alexandria, Department of Geography, ...\\
       %%      \email{c.ptolemy@hipparch.uheaven.space}
          %   \thanks{The university of heaven temporarily does not
          %           accept e-mails}
             }

   \date{Received ; accepted }

% \abstract{}{}{}{}{} 
% 5 {} token are mandatory
% \linenumbers*[1]
  \abstract
   {Regarding tidal dissipation in Saturn, usually parameterized by Saturn's quality factor Q, there remains a discrepancy between conventional estimates and the latest determination that has been derived from astrometric observations of Saturn's inner satellites. If dissipation in Saturn is as large as the astrometric observations suggest, and is independent of time and tidal frequency, conventional models predict that Mimas' initial orbit should be located inside Saturn's synchronous orbit or even inside its Roche limit, in contradiction to formation models.}
   {Using simple structure models and assuming Saturn's core to be viscoelastic, we look for dissipation models which are consistent with both the latest astrometric observations and with Mimas' orbital migration.}
   {Firstly, using a two-layer model of Saturn's interior structure, we constrain the ranges of rigidity and viscosity of Saturn's core which are consistent with Saturn's dissipation derived from astrometric observations at the tidal frequencies of Enceladus, Tethys, and Dione. Next, within the constrained viscosity and rigidity ranges, we calculate Mimas' semi-major axis considering the frequency dependence of viscoelastic dissipation in Saturn's core. By the two calculations, we evaluate (1) Saturnian models which can explain the astrometrically determined Saturnian dissipation, and (2) whether Mimas' initial semi-major axis is larger than the synchronous orbit.}
   {We show that if the core is assumed to be solid with a viscosity of 10$^{13}$-10$^{14}$ Pa s (depending on its size), the lower boundary of the observed Saturnian dissipation at tidal frequencies of Enceladus, Tethys, and Dione ($k_{2s}/Q_s$$\sim$4$\times$10$^{-5}$ where $k_{2s}$ is Saturn's second degree Love number and $Q_s$ its quality factor) can be explained by our model. In this viscosity range, Mimas can stay outside the synchronous orbit and the Roche limit for 4.5 billion years of evolution. }
   {In the case of a frequency dependent viscoelastic dissipative core, the lower boundary of the observed Saturnian dissipation can be consistent with the orbital expansion of Mimas. In this model, the assumption of a late formation of Mimas, discussed recently, is not required.}


  \maketitle
%
%-------------------------------------------------------------------

\section{Introduction}
Tidal dissipation in Saturn induced by its moons can be one important factor constraining the interior structure and dynamics of Saturn. The magnitude of dissipation is estimated by the quality factor, Q, which is defined as the ratio between the peak of the stored energy and the dissipated energy over one tidal cycle. A small Q value means that a large quantity of energy is dissipated in Saturn. 

Although the Saturnian Q value is not constrained well, its minimum value has been evaluated by the orbital expansion of Mimas \citep[e.g.,][]{goldreich1966,gavrilov1977,murray2000}. The tidal bulge on Saturn exerted by Mimas is ahead of the Saturn-Mimas axis because Saturn's rotation period is smaller than the orbital period of Mimas. This tidal bulge adds a torque to Mimas and the semi-major axis of Mimas increases. Ignoring dissipation in Mimas, the increasing rate of semi-major axis $a$ of Mimas is given \citep[e.g.,][]{murray2000} by
\begin{equation}
\frac{da}{dt}=3\sqrt{\frac{G}{M_s}}\frac{M_{m}R_s^5}{a^{5.5}}\frac{k_{2s}}{Q_s},
\label{a}
\end{equation}
where $t$, $G$, $M_s$, $M_m$ , and $R_s$ are time, the gravitational constant, Saturnian mass, Mimas' mass, and Saturnian radius, respectively. The parameters $k_{2s}$ and $Q_s$ are the second degree Love number and the Q value of Saturn. By integrating Eq. (\ref{a}) backwards in time, the past semi-major axis of Mimas can be calculated. As this equation shows, the change of the orbit becomes large when $Q_s$ is small. In order to move to the current orbit from a small initial semi-major axis in the past, Mimas must be located outside of the surface (and the Roche limit) of Saturn when it was formed. One of the simplest assumptions for the evaluation of $Q_s$ is that it is constant along the evolution of the planet and that it is independent of the forcing frequency \citep[e.g.,][]{macdonald1964,kaula1964}. The tidal Q of Saturn had so far been constrained by considering that Mimas took at least 4.5 billion years to move from a quasi-synchronous orbit with Saturn to its present location, yielding $Q_s$>18000 \citep[e.g.,][]{goldreich1966,gavrilov1977,murray2000,meyer2007}.
However, recent astrometric measurements of Saturnian satellites have indicated that the Q value of Saturn is much lower \citep{lainey2012,lainey2017}. The studies reported above discuss the Saturnian dissipation by the value of $Q_s$ assuming the value of $k_{2s}$ derived by \citet{gavrilov1977}. However, from the astrometric measurements the value of $k_{2s}/Q_s$ can be obtained rather than $Q_s$. \citet{lainey2017} show that $k_{2s}/Q_s$ of Saturn is around 10$^{-4}$ (the detailed values are shown below), which implies that $Q_s$ is a few thousand only, even if we consider the latest determination of $k_{2s}$=0.413 \citep{wahl2017}. As a mechanism of the observed dissipation of Saturn, \citet{lainey2017} suggest a viscoelastic response of the solid core. Although a viscoelastic core can generate significant dissipation \citep{remus2012,remus2015,lainey2017}, it contradicts the conventional estimations derived from the orbital expansion of Mimas \citep[e.g.,][]{goldreich1966,gavrilov1977,murray2000,meyer2007}. One hypothesis is that the Saturnian satellites such as Mimas were formed recently \citep{charnoz2011}. If Mimas moved to the current position on a small timescale, strong constant dissipation inside Saturn does not contradict the orbital expansion of Mimas. 

Dissipation of Saturn plays an important role for the possible activity of the Saturnian satellites. The Cassini probe has observed that Enceladus is a currently active body emanating water plumes, possibly implying liquid water in its subsurface \citep[e.g.,][]{porco2006}. If Enceladus is in an equilibrium state of the resonance with Dione, the tidal heating rate generated in Enceladus is $\sim$1.1 GW at $Q_s$=18,000 \citep{meyer2007}, which is independent of the interior structure of Enceladus. However, the equilibrium heating rate increases with decreasing $Q_s$ \citep{meyer2007,lainey2012}, which may affect Enceladus' structure and activity. 

In this work, although the detailed structure and the state of the Saturnian core (e.g., whether the core is solid or liquid) is uncertain, we reconsider the conventional estimations derived from the orbital migration of Mimas assuming that Saturn's core is solid and viscoelastic. In case of a viscoelastic response, both Love number and Q value depend on the frequency of the cyclic forcing. Because the tidal frequency on Saturn changes with Mimas' semi-major axis, $k_{2s}$ and $Q_s$ in Eq. (\ref{a}) cannot be constant. It has been demonstrated that constant Q is a strong assumption and frequency dependent Q can lead to different tidal rotational and orbital evolution \citep[e.g.,][]{efroimsky2007,auclair2014}. We calculate the past semi-major axis of Mimas taking into account the frequency dependence of the dissipation with the simple two-layer structure models by \citet{remus2012,remus2015}. Firstly we constrain the rigidity and viscosity of the solid core, which are consistent with the latest observational results by \citet{lainey2017}, and then we calculate the semi-major axis within the range of the constrained rheological parameters. Here, we consider the value of $k_{2s}/Q_s$ rather than $Q_s$ because $k_{2s}/Q_s$ is the directly obtained parameter by astrometric methods. By the two calculations, we suggest the Saturnian models which can be consistent with both Mimas' orbital evolution and the latest observational results of Saturnian dissipation. 
  
\section{Structure and rheology of Saturn}
\citet{remus2012,remus2015} have explained the large dissipation of Saturn using the two layer model with fluid envelope and viscoelastic solid core. Here we assume this simple structure model because the purpose of this work is to relate the observed dissipation to Mimas' orbital change. Mass and size of Saturn are shown in Table \ref{table}, which are consistent with \citet{remus2012,remus2015}. \citet{lainey2017} assume more detailed models with layered structure and the calculation method by \citet{tobie2005}, and mention that the viscosity range to generate the observed $Q_s$ is compatible between their complex model and the simple two layer model. In the structure models by \citet{remus2012,remus2015} and \citet{lainey2017}, the fluid envelope is assumed to be non-dissipative. Convective turbulent friction applied on fluid tidal waves \citep[e.g.,][]{ogilvie2004} and the resonant-lock mechanism \citep{fuller2016} may contribute to the dissipation in Saturn, which can compete with the viscoelastic dissipation in the solid core \citep{guenel2014}. Here we consider dissipation only in the solid core as an end-member model. The effect of dissipation in the envelope will be addressed in future studies. In the case of viscoelastic material, due to the delay of the response to the forcing because of viscous friction, the Love number becomes complex $\tilde{k}_{2s}$, which is given \citep{remus2015} by 
\begin{equation}
\tilde{k}_{2s}=\frac{3}{2}\frac{\tilde{\epsilon}+\frac{2}{3}\beta}{\alpha\tilde{\epsilon}-\beta},
\end{equation}
where
\begin{equation}
\alpha=1+\frac{5}{2}\left(\frac{\rho_c}{\rho_o}-1\right)\left(\frac{R_c}{R_s}\right)^3
\end{equation}
\begin{equation}
\beta=\frac{3}{5}\left(\frac{R_c}{R_s}\right)^2(\alpha-1)
\end{equation}
\begin{equation}
\tilde{\epsilon}=\frac{\frac{19\tilde{\mu}_c}{2\rho_cg_cR_c}+\frac{\rho_o}{\rho_c}\left(1-\frac{\rho_o}{\rho_c}\right)\left(\beta+\frac{3}{2}\right)+\left(1-\frac{\rho_o}{\rho_c}\right)}
{\left(\alpha+\frac{3}{2}\right)\frac{\rho_o}{\rho_c}\left(1-\frac{\rho_o}{\rho_c}\right)}.
\end{equation}
The parameters $R_c$, $g_c$, $\rho_c$ , and $\rho_o$ are radius of the core, gravitational acceleration at the surface of the core, density of the core, and density of the envelope, respectively. We consider the value of $R_c$ between 0.2$R_s$ and 0.24$R_s$, which includes the core radius by \citet{remus2015}. $g_c$, $\rho_c$ , and $\rho_o$ can be calculated from the mass of Saturn $M_s$, the mass of the core $M_c$, $R_s$ and $R_c$ shown in Table \ref{table}. $\tilde{\mu}_c$ is the complex shear modulus of the viscoelastic core. In this work, we assume a Maxwell rheology in which the complex shear modulus of the Maxwell model is given by
\begin{equation}
\tilde{\mu}_c=\frac{\omega_f\mu\eta}{\omega_f\eta+i\mu},
\label{shear_modulus}
\end{equation}
where $i=\sqrt{-1}$. The parameters $\mu$, $\eta$ and $\omega$ are the tidally effective rigidity and viscosity of the solid core of Saturn and the forcing frequency, respectively. For the frequency of tides on Saturn by satellites, $\omega_f$ can be represented by the rotational angular velocity of Saturn $\Omega$ and the mean motion of the satellites $\omega$ as $\omega_f$=2($\Omega$-$\omega$). Because $Q_s$=|$\tilde{k}_{2s}$|/|Im($\tilde{k}_{2s}$)| \citep[e.g.,][]{remus2012}, $k_{2s}/Q_s$=|Im($\tilde{k}_{2s}$)| with $k_{2s}$=|$\tilde{k}_{2s}$|.

 %--------------------------------------------------- One column table
\begin{table}[h]
%\centering
\caption{Physical parameters and values.}
\label{table}
\begin{tabular}{lcccc} \hline
%\begin{tabular}{lcccc} \hline
Parameter&Symbol&Value&Unit\\ \hline
Radius of Earth&$R_{\oplus}$&6.371$\times$10$^6$&m\\
Mass of Earth&$M_{\oplus}$&$5.9736\times10^{24}$&kg\\
Radius of Saturn&$R_s$&9.14$R_{\oplus}$&\\
Mass of Saturn&$M_s$&95.159$M_{\oplus}$&\\
Mass of solid core&$M_c$&18.65$M_{\oplus}$&\\
Rotation rate of Saturn&$\Omega$&1.65$\times10^{-4}$&rad s$^{-1}$\\
Mass of Mimas&$M_m$&3.7493$\times10^{19}$&kg\\
Current $a$ of Mimas&$a_0$&1.8552$\times10^8$&m\\ \hline
\end{tabular}
\end{table}
%                                                One column figure
%----------------------------------------------------------------- 
   \begin{figure*}[htbp]
   \centering
   \includegraphics[width=14.5cm]{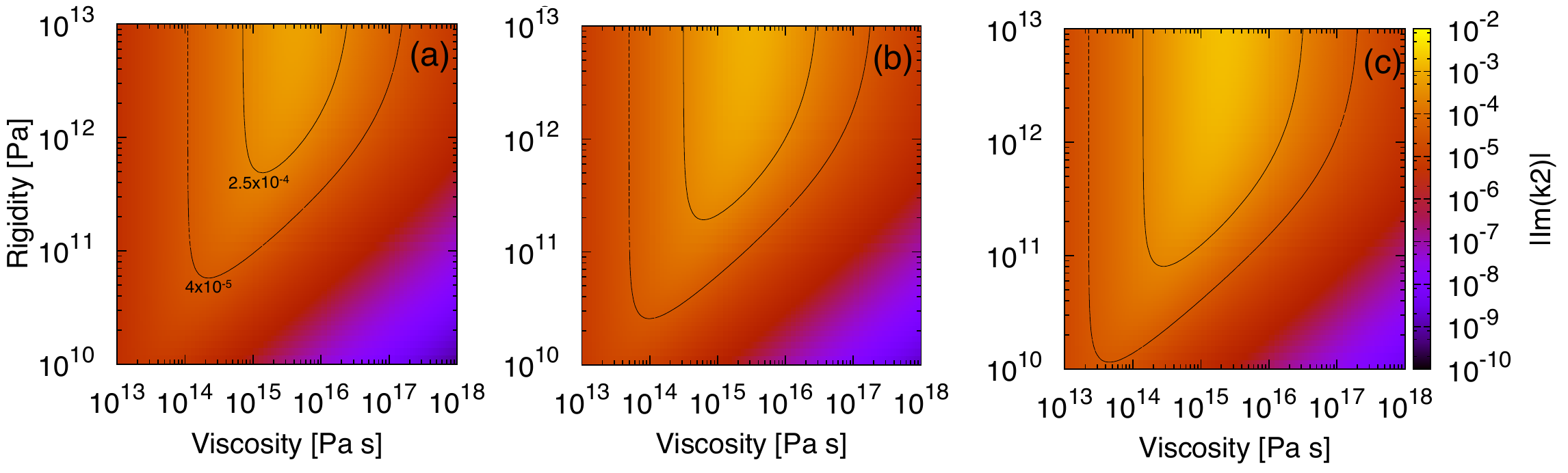}
      \caption{ |Im($\tilde{k}_{2s}$)| as a function of viscosity and rigidity of the solid core. $\omega_f$=2.5$\times$10$^{-4}$ rad s$^{-1}$ and the core radius is (a): 0.2$R_s$ (b): 0.219$R_s$ and (c): 0.24$R_s$. Two contours show the upper |Im($\tilde{k}_{2s}$)|=2.5$\times$10$^{-4}$ and the lower |Im($\tilde{k}_{2s}$)|=4$\times$10$^{-5}$ boundaries which are consistent with the observed values.}
        \label{fig1}
  \end{figure*}
%-----------------------------------------------------------------

%                                                One column figure
%----------------------------------------------------------------- 
   \begin{figure*}[htbp]
   \centering
   \includegraphics[width=15.5cm]{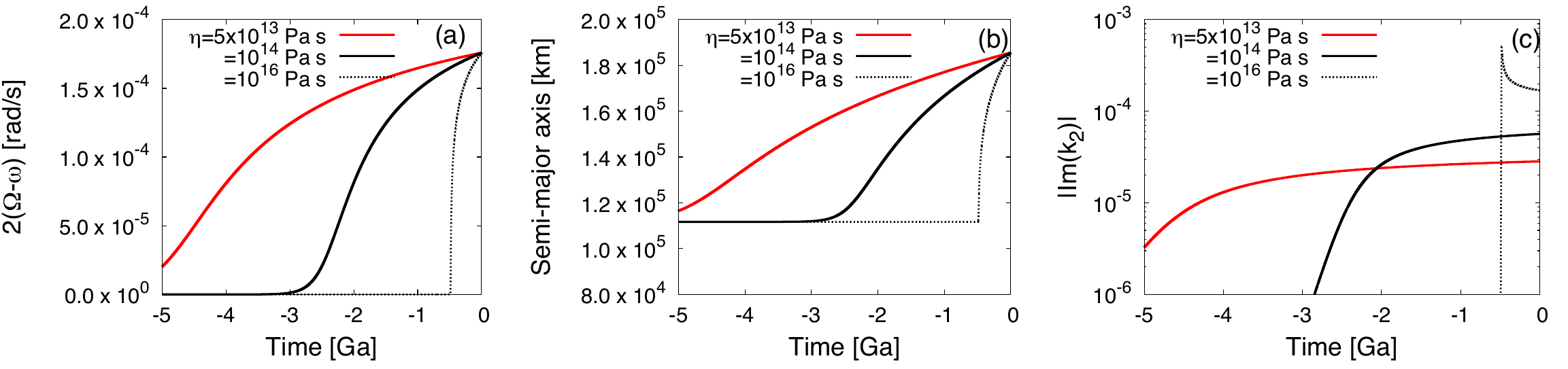}
      \caption{Evolutions of (a): 2($\Omega$-$\omega$) (b): |Im($\tilde{k}_{2s}$)| and (c): $a$ with different $\eta$ at $R_c$=0.219$R_s$ and $\mu$=5$\times$10$^{11}$ Pa. Integrations are conducted backwards in time and the current $a$ of Mimas is shown at $t$=0.}
        \label{fig2}
  \end{figure*}
%-----------------------------------------------------------------

\section{Method}
\subsection{Constraint of rigidity and viscosity for observed dissipation}
Firstly we calculate the ranges of rigidity and viscosity which are consistent with the observed $k_{2s}/Q_s$ at the tidal frequencies caused by Enceladus, Tethys, and Dione. \citet{lainey2017} also estimated $k_{2s}/Q_s$ at Rhea's frequency. However, they suggest that the dissipation at Rhea's frequency is caused by turbulent friction in the envelope, which is beyond the scope of this work. The value of $k_{2s}/Q_s$ is (20.70$\pm$19.91)$\times$10$^{-5}$, (15.84$\pm$12.26)$\times$10$^{-5}$ and (16.02$\pm$12.72)$\times$10$^{-5}$ at the tidal frequencies of Enceladus, Tethys, and Dione, respectively, and (1.59$\pm$0.74)$\times$10$^{-4}$ in a global estimation \citep{lainey2017}. Thus, we consider |Im($\tilde{k}_{2s}$)| between 4$\times$10$^{-5}$ and 2.5$\times$10$^{-4}$. $k_{2s}/Q_s$ is almost independent of the mean motion difference among these three satellites \citep{lainey2017}. We set $\omega_f$=2.5$\times$10$^{-4}$ rad s$^{-1}$, which is between the tidal frequencies of Enceladus and Dione. %We changed the value to $\omega_f$ considering the orbital frequency of Enceladus and Dione. However, as \citet{lainey2017} mention, the distribution of |Im($\tilde{k}_{2s}$)| little changes in this frequency range.

\subsection{Change of Mimas' semi-major axis}
As a next step, we calculate the initial semi-major axis of Mimas with the constrained rigidity and viscosity. Assuming a Kepler orbit, the mean motion of Mimas is given by $\omega$=$\sqrt{GM_s/a^3}$. Substituting |Im($\tilde{k}_{2s}$)| in Eq. (\ref{a}), $da/dt$ and dissipation of Saturn can be coupled by Eqs. (\ref{a})-(\ref{shear_modulus}). Integrations of $a$ are performed backwards from the current semi-major axis $a_0$=1.8552$\times$10$^8$ m  \citep{murray2000} with a Runge-Kutta method and 10$^4$ yr of time step. We performed additional calculations with time steps of 10$^3$ yr and 5$\times$10$^{2}$ yr and the results did not change. For simplicity, the angular velocity of Saturn's rotation and the mass of Mimas are fixed at $\Omega$=1.65$\times$10$^{-4}$ rad s$^{-1}$ \citep{giampieri2006,anderson2007} and $M_m$=3.7493$\times$10$^{19}$ kg \citep{jacobson2006}, respectively.

\section{Results}
\subsection{Rigidity and viscosity ranges for observed dissipation}
Figure \ref{fig1} shows |Im($\tilde{k}_{2s}$)| as a function of rigidity and viscosity of Saturn's core. The rigidity and viscosity ranges which are consistent with the observed Saturnian dissipation (Fig. \ref{fig2}) are compatible with the rheological values estimated by \citet{remus2015}. |Im($\tilde{k}_{2s}$)| becomes maximum at around 10$^{15}$ Pa s. Although rigidity and viscosity which attain the observed |Im($\tilde{k}_{2s}$)| change with the core radius, minimum viscosities for the observed dissipation should be 10$^{13}$-10$^{14}$ Pa s. If $\eta$$\sim$10$^{15}$ Pa s, |Im($\tilde{k}_{2s}$)| becomes too high to be consistent with the observed values.

\subsection{Evolutions of dissipation and Mimas' orbit}
Figure \ref{fig2} shows the evolutions of tidal frequency 2($\Omega$-$\omega$), |Im($\tilde{k}_{2s}$)| , and semi-major axis $a$ at $R_c$=0.219$R_s$. The rigidity $\mu$ is 5$\times$10$^{11}$ Pa and the viscosity $\eta$=5$\times$10$^{13}$, 10$^{14}$ and 10$^{16}$ Pa s, which are consistent with the reasonable parameter range \citep{remus2015} and the observed $k_{2s}/Q_s$ at the frequency of Enceladus, Tethys, and Dione (Fig. \ref{fig1}). Due to the dissipation of Saturn, the semi-major axis decreases with decreasing time (Mimas' orbit expands with time). Because we assume $\omega$=$\sqrt{GM_s/a^3}$, 2($\Omega$-$\omega$) also decreases as time decreases. However, once the semi-major axis decreases to the synchronous orbit with Saturn in which Mimas' orbital period is the same as the rotational period of Saturn ($\Omega$=$\omega$), dissipation does not occur because 2($\Omega$-$\omega$) becomes zero, and thus the migration of Mimas stops. At $\eta$=5$\times$10$^{13}$ Pa s and 10$^{14}$ Pa s, |Im($\tilde{k}_{2s}$)| decreases with decreasing time (Fig. \ref{fig2} b). In this viscosity range, 2($\Omega$-$\omega$)/2$\pi$ is smaller than the Maxwell frequency ($\mu/\eta$). Thus the response of the solid core is like a fluid as $a$ decreases (2($\Omega$-$\omega$) decreases) and the core becomes less dissipative. On the other hand, at $\eta$=10$^{16}$ Pa s, |Im($\tilde{k}_{2s}$)| increases with decreasing time because 2($\Omega$-$\omega$)/2$\pi$ at $t$=0 is larger than the Maxwell frequency. By the decrease of 2($\Omega$-$\omega$) (increase of $\omega$), the response of the solid core becomes viscoelastic from elastic, which results in large dissipation. Due to the large dissipation, $a$ gets into the synchronous orbit at around -0.5 Ga (Fig. \ref{fig2} c). 

%                                                One column figure
%----------------------------------------------------------------- 
   \begin{figure}[htbp]
   \centering
   \includegraphics[width=5cm]{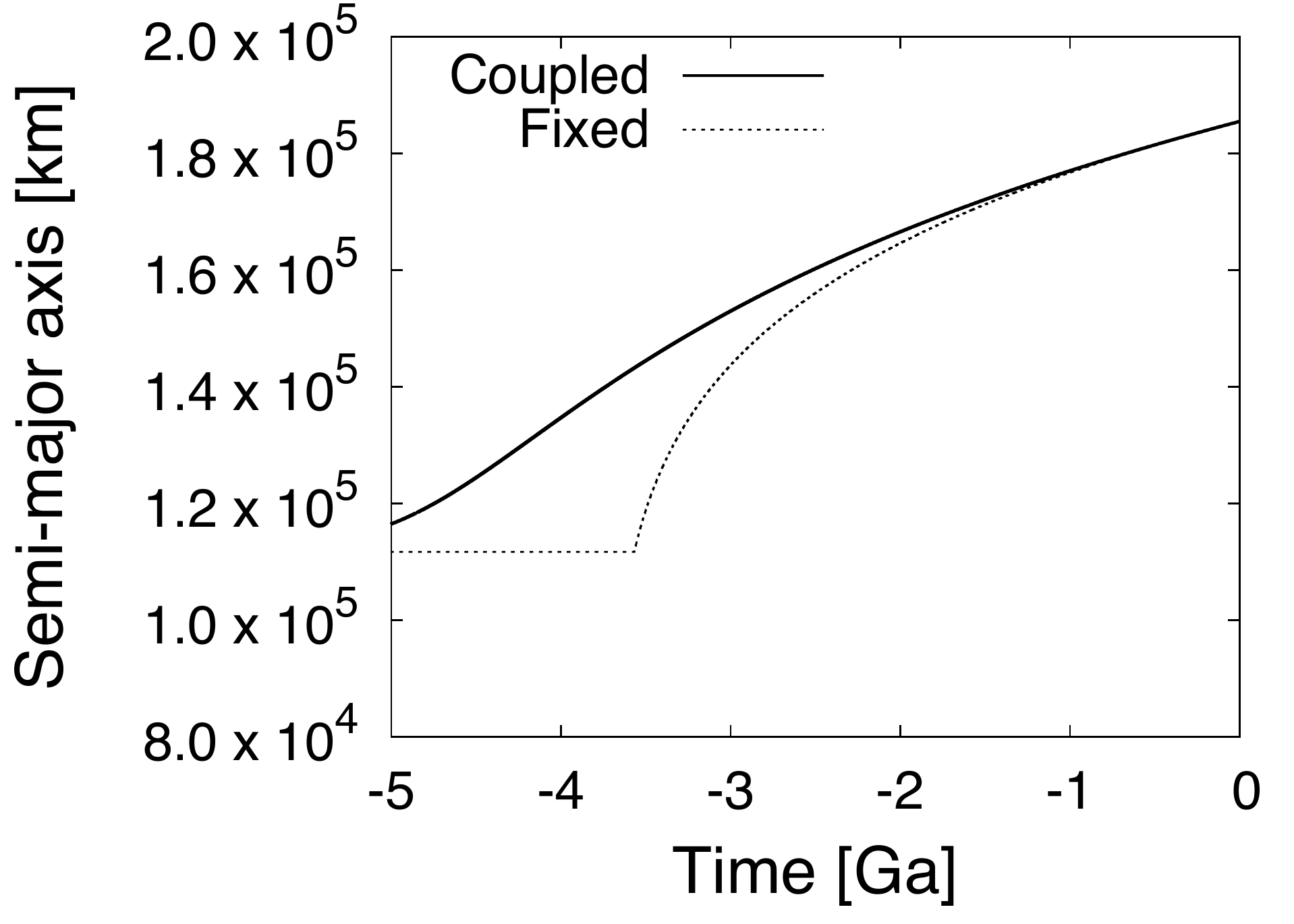}
      \caption{Comparison of semi-major axis between constant |Im($\tilde{k}_{2s}$)| at 2.8$\times$10$^{-5}$ and coupled orbital |Im($\tilde{k}_{2s}$)| at $R_c$=0.219$R_s$ and $\mu$=5$\times$10$^{11}$ Pa. Once Mimas gets into the synchronous orbit, we assume that $a$ does not change.}
        \label{fig3}
  \end{figure}
%-----------------------------------------------------------------

One important result is that, at $\eta$=5$\times$10$^{13}$ Pa s, 2($\Omega$-$\omega$) does not become zero and thus $a$ does not get into the synchronous orbit at the time of solar system formation (around -4.5 Ga). Thus, if the viscosity of Saturn is around $\eta$=5$\times$10$^{13}$ Pa s, the latest observational dissipation at the frequency of Enceladus, Tethys, and Dione does not contradict the conventional evaluations that Mimas must have been outside of the surface of Saturn and the synchronous orbit 4.5 billion years ago \citep[e.g.,][]{goldreich1966,gavrilov1977,murray2000,meyer2007}. In addition to the synchronous orbit, Mimas should be outside of the Roche limit. Assuming that Mimas is a rigid and spherical body, the distance of the Roche limit $a_L$=$R_s$(3$\rho_s/\rho_m$)$^{1/3}$ \citep{murray2000}, while $a_L$=2.456$R_s$($\rho_s/\rho_m$)$^{1/3}$ if Mimas is fluid \citep{chandra1969}, where $\rho_s$ and $\rho_m$ are the density of Saturn and Mimas, respectively. Using the values of our model ($\rho_s$=687.3 kg m$^{-3}$) and the mean density of Mimas at $\rho_m$=1150 kg m$^{-3}$ \citep{thomas2007}, $a_L$ is $\sim$7$\times$10$^4$ km for the rigid Mimas and $\sim$1.2$\times$10$^{5}$ km in the case of the fluid Mimas, respectively. Even if Mimas is a fluid body, it stays outside of the Roche limit at $\eta$=5$\times$10$^{13}$ Pa s (Fig. \ref{fig2}). 

We calculated Mimas' orbit with different rigidity within the consistent range to the observations. Although the magnitude of dissipation depends on rigidity, the value of viscosity strongly affects whether Mimas gets into the synchronous orbit because |Im($\tilde{k}_{2s}$)| is relatively independent of rigidity at $\eta\sim$10$^{13}$-10$^{14}$ Pa s (Fig. \ref{fig1}). Thus, as long as the rigidity is consistent with the observations (Fig. \ref{fig1}), the conclusion that Mimas does not get into the synchronous orbit at the lower boundary of the viscosity does not change.

In addition to low viscosity, the frequency dependence of |Im($\tilde{k}_{2s}$)| is another reason why Mimas does not get into the synchronous orbit and the Roche limit. In the case of $\eta$=5$\times$10$^{13}$ Pa s, |Im($\tilde{k}_{2s}$)| is approximately 2.8$\times$10$^{-5}$ at $t$=0 (Fig. \ref{fig2} b). Figure \ref{fig3} compares $a$ of the case in which |Im($\tilde{k}_{2s}$)| is fixed at 2.8$\times$10$^{-5}$ with the frequency dependent |Im($\tilde{k}_{2s}$)| (coupled model of dissipation and orbit). While the semi-major axis is larger than the synchronous orbit for 4.5 billion years in the coupled model, Mimas' orbit becomes synchronous with Saturn at -3.5 Ga in the case of fixed |Im($\tilde{k}_2$)|. As the semi-major axis decreases, due to the fluid response, |Im($\tilde{k}_{2s}$)| decreases (Fig. \ref{fig2} b) in the coupled model. Thus, $da/dt$ becomes smaller as compared to the case with fixed |Im($\tilde{k}_{2s}$)|, which results in the slow migration of Mimas.

\begin{figure}[htbp]
   \centering
   \includegraphics[width=9.5cm]{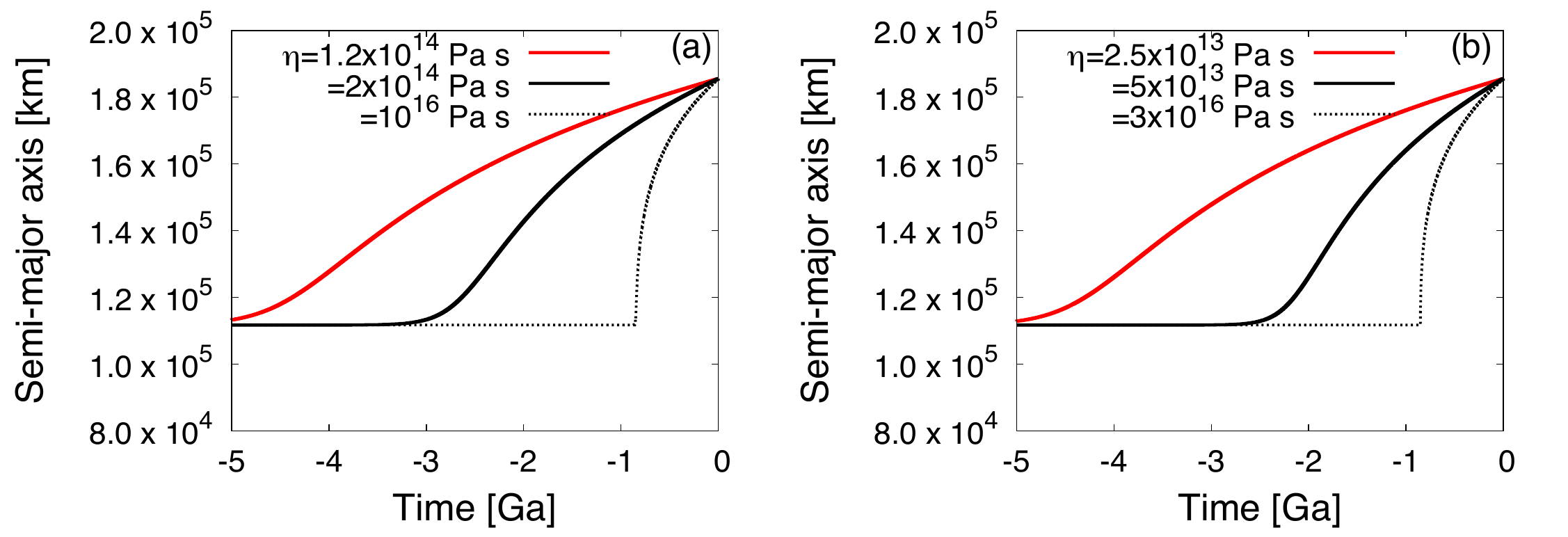}
      \caption{Evolutions of semi-major axis at (a): $R_c$=0.2$R_s$ and (b): $R_c$=0.24$R_s$ and $\mu$=5$\times$10$^{11}$ Pa and viscosities consistent with the observed dissipation.}
        \label{fig4}
  \end{figure}

Changes of semi-major axis at $R_c$=0.2$R_s$ and 0.24$R_s$ are shown in Fig. \ref{fig4}, respectively. Because the changing rate of the semi-major axis depends on |Im($\tilde{k}_{2s}$)|, Mimas can avoid getting into the synchronous orbit regardless of core radius if the viscosity of Saturn's core is a lower boundary for the observed $k_{2s}/Q_s$. It is uncertain whether the viscosity of Saturn's solid core could be in the order of 10$^{13}$ Pa s. If 10$^{14}$ Pa s of viscosity is required, Saturn should have a small core at around 0.2$R_s$. Future studies with ab-initio calculations of the equation of state (EOS) and observations can constrain the interior structure of Saturn more precisely \citep[e.g.,][]{helled2013,miguel2016}.

%                                                One column figure
%----------------------------------------------------------------- 
%-----------------------------------------------------------------

One caveat of our results is that, although $\sim$4$\times10^{-5}$ of |Im($\tilde{k}_{2s}$)| is consistent with the observational values derived from each Saturnian satellite, it is smaller than the comprehensive evaluation at |Im($\tilde{k}_{2s}$)|=(1.59$\pm$0.74)$\times$10$^{-4}$ \citep{lainey2017}. In the case of the simple two layer model, if |Im($\tilde{k}_{2s}$)|$\sim$8$\times10^{-5}$, Mimas gets into the synchronous orbit from the current position $\sim$3 billion years ago. Calculations with more accurate core structure may solve this discrepancy, which will be addressed in future studies. 

If we apply Eq. (\ref{a}) to Enceladus, Tethys, and Dione, $da/dt$ of Tethys with the current parameters could be larger than that of Mimas. Thus Tethys can expand faster than other satellites and further discussions about Tethys' orbital evolution may be required in future studies. The fast expansion is due to Tethys' greater mass (as compared to Mimas and Enceladus) and its relatively small semi-major axis (as compared to Dione). It is indicated that Tethys experienced large tidal heating like Enceladus in the past \citep{giese2007,chen2008}. Thus, the effect of dissipation in Tethys may not be negligible for its orbital evolution, which modifies Eq. (\ref{a}) and requires detailed evolutions of the eccentricity and interior structure of Tethys.

\section{Conclusions}
Conventionally, the magnitude of Saturnian dissipation has been constrained from the orbital expansion of Mimas. Assuming a constant Love number, more than a few tens of thousands of Saturnian Q is required for Mimas to stay outside of the surface of Saturn or the synchronous orbit 4.5 billion years ago \citep[e.g.,][]{goldreich1966,gavrilov1977,murray2000,meyer2007}. However, the latest observations of Saturnian dissipation estimated at the orbital frequency of Enceladus, Tethys, and Dione imply that the Saturnian Q is in the order of a few thousand, only \citep{lainey2017}. We calculated the past semi-major axis of Mimas induced by tidal dissipation in Saturn's solid core assuming a Maxwell rheology and frequency dependence of $k_{2s}/Q_s$ (|Im($\tilde{k}_{2s}$)|). If the viscosity of Saturn's core is consistent with the lower boundary of the observed dissipation ($k_{2s}/Q_s\sim4\times10^{-5}$) \citep{lainey2017} and $\mu/\eta$ is larger than 2($\Omega$-$\omega$)/2$\pi$, due to the smaller tidal frequency in the past, Mimas can stay well outside the synchronous orbit even if it  formed 4.5 billion years ago. The viscosity consistent with the observations and Mimas' expansion changes with the radius of the Saturnian core. If the core radius is 0.2$R_s$, 0.219$R_s$ and 0.24$R_s$, $\sim$10$^{14}$ Pa s, $\sim$5$\times$10$^{13}$ Pa s, and $\sim$2.5$\times$10$^{13}$ Pa s of viscosities are required, respectively. In the case of these viscosity values, dissipation of Saturn is consistent with both the latest observational results and Mimas' orbital evolution. In this model, the assumption of a late formation discussed recently \citep{charnoz2011} is not required.

In this work, we assume that the solid core and the envelope are homogeneous. In addition, the envelope is assumed to be non-dissipative, and rigidity and viscosity of the core do not change with time. So far, whether a solid state core exists and its detailed structure are uncertain. We note that the Maxwell rheology used in our model would not be applicable if Saturn has a liquid core. Detailed interior structure models of Saturn and thermal evolution models can constrain the dissipation mechanisms more precisely, which should be addressed in future studies.

\begin{acknowledgements}
This work was supported by a JSPS Research Fellowship.
\end{acknowledgements}


\end{document}